\documentclass[APA,Times2COL,colorlinks=true]{WileyNJDv5} 
\copyyear{2023}
\startpage{1}
\hypersetup{linkcolor=blue,citecolor=blue}
\usepackage{color}
\usepackage{soul}

\begin{document}

\title{Adaptive sampling method to monitor low-risk pathways with limited surveillance resources}

\author[1,2,3]{Thao P. Le}
\author[1,2,3]{Thomas K. Waring}
\author[2,3]{Howard Bondell}
\author[1]{Andrew P. Robinson}
\author[1,2,3]{Christopher M. Baker}

\authormark{LE \textsc{et al.}}
\titlemark{ADAPTIVE SAMPLING METHOD TO MONITOR LOW-RISK PATHWAYS WITH LIMITED SURVEILLANCE RESOURCES}

\address[1]{\orgdiv{The Centre of Excellence for Biosecurity Risk Analysis}, \orgname{The University of Melbourne}, \orgaddress{\state{Victoria}, \country{Australia}}}
\address[2]{\orgdiv{Melbourne Centre for Data Science}, \orgname{The University of Melbourne}, \orgaddress{\state{Victoria}, \country{Australia}}}
\address[3]{\orgdiv{School of Mathematics and Statistics}, \orgname{The University of Melbourne}, \orgaddress{\state{Victoria}, \country{Australia}}}

\abstract[Abstract]{
The rise of globalisation has led to a sharp increase in international trade, with high volumes of containers, goods and items moving across the world. Unfortunately, these trade pathways also facilitate the movement of unwanted pests, weeds, diseases, and pathogens. Each item could contain biosecurity risk material, but it is impractical to inspect every item. Instead, inspection efforts typically focus on high risk items. However, low risk does not imply no risk. It is crucial to monitor the low risk pathways to ensure that they are and remain low risk. To do so, many approaches would seek to estimate the risk to some precision, but the lower the risk, the more samples needed to estimate the risk. On a low-risk pathway that can be afforded more limited inspection resources, it makes more sense to assign fewer samples to the lower risk activities.  We approach the problem by introducing two thresholds. Our method focuses on letting us know whether the risk is below certain thresholds, rather than estimating the risk precisely. This method also allows us to detect a significant change in risk. Our approach typically requires less sampling than previous methods, while still providing evidence to regulators to help them efficiently and effectively allocate inspection effort.}


\maketitle

\section{Introduction}\label{sec:Introduction}

International trade increases the likelihood of new pest incursions, and these risks are increasing over time. According to the World Trade Organisation, world trade volume has increased nearly 45 times since 1950 \cite{WTO_stats}. In 2021 alone, international merchandise exports were worth US\$22.3 trillion, and over 11.0 billion tons of goods were loaded for international maritime trade \cite{UNCTAD_stats_2022}. These high-volume trade pathways facilitate the movement of unwanted, invasive and non-indigenous pests, weeds, diseases, and pathogens. These pests can disperse into new environments at an unprecedented rate along these trade pathways, be it accidental or deliberate, and cause significant ecological, economic, and human health consequences.

The economic impacts from biological invasions are significant. The costs of multiple biological invasions, their impact, and their management, has been estimated to cost the United States over USD\$21 billion annually over 2010-2020 \cite{FantleLepczyk2022}, and estimated to cost a minimum of USD\$1.288 trillion worldwide over 1970--2017 \cite{Diagne2021}. One example of a major pest is the brown marmorated stink bug (BMSB), which has spread into North America and Europe and caused significant economic impacts \cite{Leskey2018}. BMSB has caused estimated crop losses of 20\% to over 90\% \cite{Leskey2010} and millions of dollars of damage to mid-Atlantic apple crops in 2010 \cite{Herrick2011}. Meanwhile, emerald ash borers and other invasive insects are estimated to kill 1.4 million street trees in the United States between 2020--2050, costing over US\$900 million for replacement \cite{Hudgins2022}.

Inspection and intervention at the border are one of the key strategies to prevent biological incursions and spread. For example, New Zealand regularly intercepts live and dead BMSB at the border each year \cite{BMSB_NZ}, including 61 live BMSB during 2021/22, and as of 2022, there is no evidence of any established BMSB population \cite{MPI_BMSB}. Aside from stopping the incursion of biological pests and reducing invasion risk, inspections and surveillance can also give us information about pest arrival rates and composition \cite{Turner2021}, thus allowing further targeting of inspection and other biosecurity strategies. Inspections can also encourage importer compliance \cite{Hester2020} by streamlining the import processes of compliant importers. 

However, surveillance systems require resources and prioritisation. Given the sheer volume of international trade and movement, it is infeasible to inspect every single parcel, container, or item that passes through a country border, due to both cost and delays in trade movement. Effective systems must prioritise within their budget constraints, while ideally achieving its intended purpose within some level of tolerance. Often, the priority is to target high-risk pathways and reduce invasion risk to a pre-defined acceptable level \cite{WTOALOP}. 

An important aspect of designing an effective surveillance system is knowing the underlying rates at which incoming items are contaminated or non-compliant, and where the pests with high possible impact and damage are likely to come from, and then applying differentiated surveillance and controls based on stratified risk---i.e., high risk and low risk \cite{Powell2015, Turner2020, Saccaggi2022}. Inspection data and other known information about pest distribution and pest impact are crucial to determining which pathways are high-risk and should be targeted for border inspections. Based on 1.9 million border interception records, \citeA{FennMoltu2022} evaluated that high-risk pathways for insect pests include plants, wood, textiles, animal products, food stuffs, live plants and cut flowers, fruit and nuts, and items with wood packaging. \citeA{Kriticos2017} examined the possible future distribution of BMSB using bioclimatic niche model, thus determining which locations are at risk of BMSB establishment. Meanwhile, \citeA{Saccaggi2022} used machine-learning on inspection data to determine the strongest predictors of pest interception and found that factors such as country, crop, and year of import can help predict future interceptions.

Given that risk is the combination of likelihood and consequence, this often involves focusing inspection resources on pathways that have either high infestation, or have damaging pests, or both. For example, Australia has a ``Country Action List (CAL)'', defining high-risk pathways for sea containers and breakbulk cargo based on documented high contamination or have high risk pests. These have mandatory inspection \cite{CAL_containers}. \citeA{Turner2020} developed a arrival-interception-establishment model combined with insect interception data to predict the number of invasive pests that were not detected by biosecurity and could establish.

While there has been a lot of work on profiling high-risk pests and items, and while border inspections target these high-risk pathways, there is less work looking at the rest of the low-risk pathways. Low-risk pathways still need to monitored: compliance surveys and/or general random surveys should be conducted on low-risk pathways so that we can ensure the risk remains low, and to know whether it is no longer low risk. Monitoring these low-risk pathways should be an important part of overall trade pathway management and analysis.

However, there is an additional challenge: the majority (or all) of the resources were concentrated on the high-risk pathways for a reason, and low-risk pathways can be very large. Our problem is thus: given limited resources for surveillance, what is the best way to sample such that we have sufficient information regarding the contamination prevalence, on a low-risk pathway?

There are existing methods for setting sample sizes, but these were not designed for low-risk pathways and/or do not allow for time-varying sampling. Sampling methods that suggest high volumes of inspection (e.g., 20\% of all items) are not feasible, and perhaps ineffective if the goal is to \emph{monitor} the low-risk pathway, rather than to prevent all possible (ideally low-risk) incursions. \citeA{Robinson2011} have a sample-size-selection method that uses the null hypothesis that the contamination rate is above a specified level based on a risk cutoff, and returns the volume that should be sampled by using a power analysis. However, the method can lead to excessively high sample sizes as the estimated rate approaches the cutoff.

The simplest method is to choose a fixed number, which we could chose to be small. For example, Australian phytosanitary inspection can involve sampling 600 units in a consignment \cite{AusHorticulture}, as this will result in a 95\% chance of detecting contamination if the infestation rate is 0.5\% \cite{ISPM_31,Whyte2009_ISPM}. In the low-risk setting, we could apply a 600-sampling volume across the entire low-risk pathway. However, this method does not vary through time and thus cannot adjust to changing risk situations.

In this paper, we formulate a new sampling method that identifies the sampling effort required to be confident that the leakage rate is below a set threshold. This method uses past data of sampling effort and interceptions to recommend minimum future sample sizes. We start by describing our new methodology, and we compare the performance of our method to \citeA{Robinson2011}'s risk-sensitive method and the fixed 600-volume method via simulation. We find that our model is able to  detect upward changes in contamination rates in a timely manner, with reduced sampling compared to \citeA{Robinson2011} and will adjust to be less effort than 600 if there is sufficient evidence that the pathway is very low risk. In general, our method can be used in any situation where assurance is sought about whether some pathway or stream of entities has rates of non-compliance between some low but nonzero risk threshold---but where knowing the exact non-compliance rate is not important---and where the non-compliance rate of the items are expected to be related by some underlying factor(s).

In Section~\ref{sec:Method}, we present our new survey design and inspection method. In Section~\ref{sec:Scenario-examples}, we compare the performance of the three different sampling methods on simulated scenarios. We discuss the results and our method in Section~\ref{sec:Discussion} and conclude in Section~\ref{sec:Conclusion}.

\section{Method}\label{sec:Method}

We are addressing the following ongoing surveillance problem: how many items should we inspect on a low risk-pathway, each reporting period, to be confident that the low-risk items are indeed low risk? The solution should allow us to sample less when we believe the risk is very low, and the solution should be able to detect when our risk is rising.

Our method uses two thresholds---one to detect changes in risk, and one to signal when the pathway is no longer low risk. 

\subsection{Conceptual model of low-risk pathways}

We consider a pathway to be a series of incoming items that can be inspected, such as containers or goods. The pathway may involve only a certain type of container or cargo, or involve only certain transport routes, or importers, or shippers, or source countries etc., or some combination of these selection criteria. 

An item can be non-compliant if it contains something of biosecurity concern (e.g. invasive and non-indigenous pests, weeds, diseases, and pathogens). There can also be other forms of non-compliance, such as documentation-related non-compliance. In general, we say that items with non-desirable elements to be ``contaminated'' or ``non-compliant''. Contamination that is not detected and released from all interventions forms the ``leakage'' from the pathway. The true unknown leakage rate is $r_{\text{true}}$, which is the proportion of contaminated items.

Our focus is on low-risk pathways. This assumes that pathway managers have conducted risk-profiling to stratify the risk in different pathways. We also assume that we have some prior inspection data on the pathway (such as from a pilot study, past inspection regimes, or other historical data), such that the pathway is deemed low-risk according to that prior data. As part of this low-risk designation, we assume that pathway managers and policy makers have defined a low-risk threshold, $T_{\text{risk}}$, and that if contamination rises about this threshold, then the pathway would no longer be considered low-risk. We also assume that the low-risk pathway comprises a very large number of units. 

\subsection{Sampling and risk-evaluation process}

\begin{figure*}
\centerline{\includegraphics[width=\textwidth]{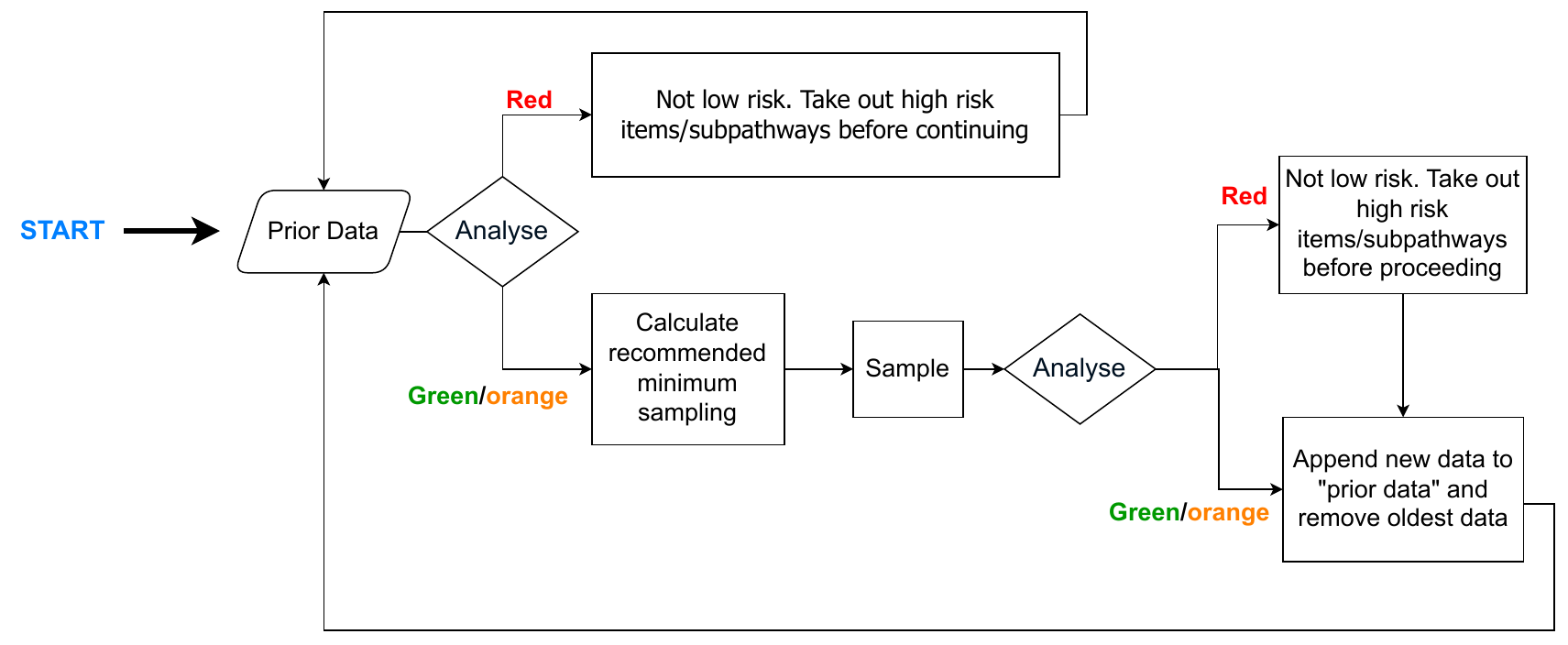}}
\caption{The overall sampling and risk evaluation process. Using prior data, we can determine the minimum recommended sampling volume for the next reporting period. After sampling, we update our belief about the leakage risk and evaluate the pathway. \label{fig:sampling_process}}
\end{figure*}

The overall process of our sampling method is shown in Figure~\ref{fig:sampling_process}. An R implementation of the  method is available.\footnote{\url{https://github.com/cebra-analytics/low_risk_sampling}}

We are interested in the leakage rate, $r_{\text{true}}$. There will always be uncertainty about the value, as we cannot sample the full pathway in our setting. However, we can use any conducted inspections to quantify our belief of the rate.

Note that our method will not be focused on estimating the true rate of contamination. Instead, we are focused on determining, with sufficient confidence, whether we are below the low-risk threshold, $T_{\text{risk}}$.

In this section, we go through the different steps in our method.

\subsubsection{Step 0: The thresholds}

Our method relies on two thresholds, the low-risk threshold $T_{\text{risk}}$, and a lower change-detection threshold, $T_{\text{change}}$. Non-compliance rates above the upper threshold, $T_{\text{risk}}$, are considered unacceptable. Meanwhile, the lower threshold $T_{\text{change}}$ is tunable, and we use this to detect changes in the non-compliance rate.

\subsubsection{Step 1: Estimating the leakage rate given past data and the colour status system\label{subsec:Estimating-leakage-rates}}

The initialisation of the problem follows \citeA{Robinson2011}.  We need to be able to mathematically encode our belief about the leakage rate, $r$ from the data. To do this, we will encode our belief of the leakage rate $r$ using a Beta probability distribution
\begin{align}
r&\sim\text{Beta}\left(\alpha, \beta\right),
\end{align}
with shape parameters $\alpha$, $\beta$.

In the absence of data, we begin with the uninformative Jeffrey's prior \cite{cai2005}, with $\alpha = \beta = 0.5$:
\begin{align}
r&\sim\text{Beta}(0.5, 0.5).
\end{align}

We assume that we have some past data about the pathway: $y_{0}$ detected contamination, among $N_{0}$ samples. This information is important, to fulfil our starting assumption that the pathway is historically low-risk. We assume that the pathway is very large, and that the number of leakages obeys a binomial distribution,
\begin{align}\label{eq:y0_binomial}
y_{0}\left|N_{0}, r \right. & \sim\text{Binomial}(N_{0},r),
\end{align}
Because we have chosen to use a Beta distribution to encode our belief about the underlying rate, $r$, we can easily update the initial uninformative prior $\text{Beta}(0.5, 0.5)$ with the past data results as follows \cite[\S2.1, \S2.8]{gelmanbda04}: 
\begin{align}
 & r\left|N_{0}, y_{0} \right.\sim\text{Beta}(\alpha_{0},\beta_{0})\nonumber \\
 & \qquad\text{where}:\;\alpha_{0}=y_{0}+0.5\\
 & \qquad\phantom{\text{where}:\;\;}\beta_{0}=N_{0}-y_{0}+0.5.\nonumber 
\end{align}
The beta distribution now encodes our belief about the true leakage rate, using the data previously gathered. Intuitively, the more parcels are inspected, the more certain we are about the closeness of the sampled leakage rate versus the underlying true leakage rate. 

Up to this point, we have followed the same steps as \citeA{Robinson2011} to assign confidence about underlying leakage rates given the available data. In the following steps, we depart from \citeA{Robinson2011}'s method with a different sample-size calculation suited for low-risk low-resourced pathways. Our method relies on a colour status system.

\begin{figure*}
\begin{centering}
\includegraphics[width=\textwidth]{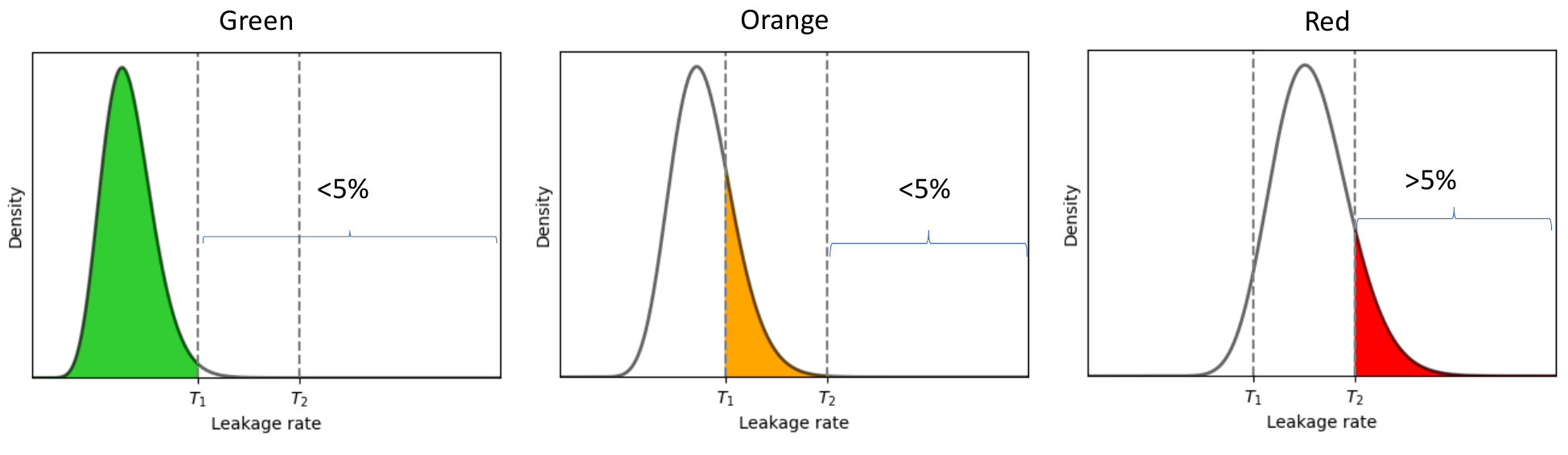}
\par\end{centering}
\caption{Our state of belief regarding the leakage rate. Green --- the 95\% credible interval is below the threshold $T_{\text{change}}$; orange --- the 95\% credible interval is below the threshold $T_{\text{risk}}$; red --- otherwise. \label{fig:colour_statuses}}
\end{figure*}

Based on the thresholds and the shape parameters $\alpha$, $\beta$, we will categorise our belief about the leakage rate using a `traffic-light' colour system, as depicted in Figure~\ref{fig:colour_statuses}:
\begin{align}
    &c\left(r\left|\alpha,\beta,T_\text{change},T_\text{risk}\right.\right) \\
    &= \begin{cases}
        \text{green,} & \text{if } \mathbb{P}\left(r<T_{\text{change}}\left| \alpha, \beta \right.\right)\ge0.95, \\
        \text{orange,} & \text{if } \mathbb{P}\left(r<T_{\text{risk}}\left| \alpha, \beta \right.\right)\ge0.95, \\
        \text{red,} & \text{otherwise.}
    \end{cases}\label{eq:colour_statues}
\end{align}
That is,
\begin{itemize}
\item The status is ``green'' if the upper one-sided 95\% credible interval of our belief is below $T_{\text{change}}$, i.e. ;
\item The status is ``orange'' if the upper one-sided 95\% credible interval of our belief is below below $T_{\text{risk}}$;
\item And the status is ``red'' otherwise. 
\end{itemize}
Note that both statuses orange and red are warnings that the risk may be rising, and are causes for concern.

\subsubsection{Step 2: Calculating the recommended minimum sampling\label{subsec:Sampling}}

For our method to work, the pathway must begin in low risk (green) status. This means that, based on the prior data $\left(N_0, y_0 \right)$, the 95\% credible interval of our belief is below $T_{\text{risk}}$, i.e. 
\begin{equation}
\mathbb{P}\left(r<T_{\text{risk}}\left| N_0, y_0 \right.\right)\ge0.95.\label{eq:low_risk_requirement}
\end{equation}
If the pathway does not satisfy this condition, then it is not low risk, and the high-risk subpathways' data should be removed before proceeding.

Now, with the low-risk pathway, we proceed by tuning the change-detection threshold $T_{\text{change}}$, such that
\begin{align}
\mathbb{P}\left(r<T_{\text{change}}\left| N_0, y_0 \right.\right) & \stackrel{!}{=}0.95,\label{eq:T_change_definition}
\end{align}
i.e., $T_{\text{change}}$ marks the one-sided 95\% credible interval of the prior beta distribution based on past data $(N_{0},y_{0})$.

Note that we could also define $T_{\text{change}}$ at a different credible interval, e.g. the 94\% or 96\% one-side credible interval, provided that the requirement in Eq.~\eqref{eq:low_risk_requirement} is still satisfied.

Now, if we take $N_1$ samples and detect $y_1$ contaminated samples, then our leakage rate posterior would be updated as follows \cite[\S2.4]{gelmanbda04}:
\begin{align}
 & r\left|N_{0}, y_{0},N_1, y_1 \right.\sim\text{Beta}(\alpha_{1},\beta_{1})\nonumber \\
 & \text{where}:\;\alpha_{1}=y_{1}+\alpha_{0}=y_1+y_0+0.5\label{eq:posterior_r}\\
 & \phantom{\text{where}:\;\;}\beta_{1}=N_{1}-y_{1}+\beta_{0}=N_1-y_1 + N_0-y_0 +0.5.\nonumber
\end{align}

Our aim is to take $N_{1}$ samples in the next reporting period, \emph{such that }our updated belief has the correct resulting colour according to the underlying situation, with some confidence. The colour assignment we want is given in Table~\ref{tab:colour_assignment_table}.

\begin{table}[!hbp]
\centering \caption{Table summarising correct posterior colour assignment $c\left(r\left|\alpha_1,\beta_1,T_\text{change}\left(N_0,y_0\right),T_\text{risk}\right.\right)$ given the true rate $r_{\text{true}}$ after updating with new data, i.e. with posterior $r\left|N_0, y_0, N_1, y_1\right.$ as in Eq.~\eqref{eq:posterior_r}. (Note that $T_{\text{change}}$ is defined relative to the prior data.)}
\label{tab:colour_assignment_table}
\centering{}%
\begin{tabular}{ll}
\toprule 
\multicolumn{1}{l}{\textbf{True rate}} & \multicolumn{1}{c}{\textbf{Wanted posterior colour assignment}}\tabularnewline
\midrule 
$r_{\text{true}}\leq T_{\text{change}}$ & Green or orange\tabularnewline
$T_{\text{change}}<r_{\text{true}}<T_{\text{risk}}$ & Green or orange or red\tabularnewline
$r_{\text{true}}\geq T_{\text{risk}}$ & Orange or red\tabularnewline
\bottomrule
\end{tabular}
\end{table}

In particular, given that we are focused on being able to detect if the pathway is increasing in risk, we calculate the minimum number of samples such that, should the true rate of non-compliance change to be above $T_{\text{risk}}$, we will obtain an orange or red posterior belief with probability $95\%$, averaged over leakage rates above $T_{\text{risk}}$ (weighted by the probability from the prior distribution).

\begin{figure*}
\begin{centering}
\includegraphics[width=0.8\textwidth]{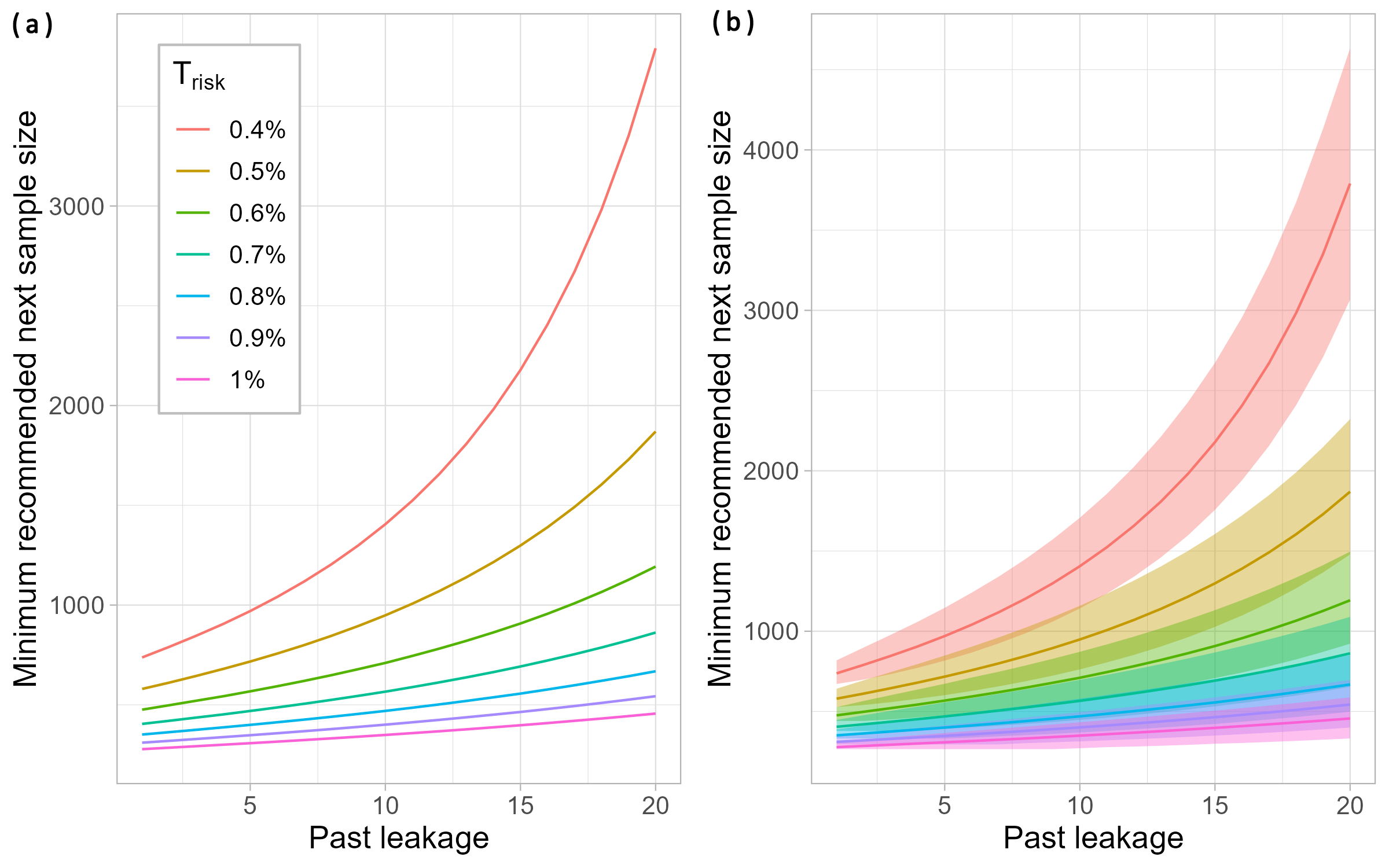}
\par\end{centering}
\caption{The calculate recommended minimum sampling volume given different choices choice of $T_{\text{risk}}$ and different past detected leakages ($y_0$). Here, we assume there were $N_0 = 10,000$ prior samples. \textbf{(a)} We choose $T_{\text{change}}$ at the upper limit of the one-sided 95\% credible interval of leakage rate belief given prior data (see Eq.~\eqref{eq:T_change_definition}) \textbf{(b)} The range is derived from different definitions of $T_{\text{change}}$ (range between 94\% and 96\% of the one-sided credible interval, where the 96\% interval requests more sampling).
\label{fig:opt_samples_plot}}
\end{figure*}

In Figure~\ref{fig:opt_samples_plot}, we have calculated the recommended sampling for different parameters. 

Mathematically, consider a fixed $N_{1}$, and a hypothetical non-compliance rate $r'>T_{\text{risk}}$. We can calculate the probability that the policy of taking $N_{1}$ samples is successful: 
\begin{align}
\mathbb{P}&\left(\text{success}\left| N_{1},r'>T_\text{risk}, N_0, y_0\right.\right) \nonumber\\
&=\sum_{y_{1}=0}^{N_{1}}\mathbb{P}\left(y_1 \left| N_{1},r'\right.\right)\cdot f\left(r\left|N_0, y_0, N_1, y_1\right.\right),
\end{align}

where $f\left(r\left|N_0, y_0, N_1, y_1\right.\right)$ flags whether the leakage rate posterior given new data $\left(N_1,y_1\right)$ is orange or red, or not:
\begin{align}
&f\left(r\left|N_0, y_0, N_1, y_1\right.\right) \nonumber\\
&=\begin{cases}
1 & \text{if }\mathbb{P}\left(r<T_\text{change}\left(N_0, y_0\right)\left|N_0, y_0, N_1 ,y_1 \right.\right)<0.95,\\ 
0 & \text{otherwise.}
\end{cases}
\end{align}

As above (Eq.~\ref{eq:y0_binomial}), the detected leakage follows a binomial distribution:
\begin{align}
y_{1}\left|N_{1}, r' \right. & \sim\text{Binomial}(N_{1},r'),
\end{align}

Our model recommends sampling $N_{1}$ times, where $N_{1}$ solves: 
\begin{align}
&\mathbb{P}\left(\text{success}\left| N_{1},N_0, y_0\right.\right)\nonumber\\
& =\int_{T_{\text{risk}}}^{1}\mathrm{d}r^\prime\,\mathbb{P}\left(\text{success}\left| N_{1},r^\prime,N_0, y_0\right.\right)\cdot \mathbb{P}\left(r^\prime\left| r^\prime>T_{\text{risk}},N_0, y_0\right.\right)\nonumber\\
 & \stackrel{!}{\geq}0.95,\label{eq:prob_correct_N1}
\end{align}
and where $\mathbb{P}\left(r^\prime\left| r^\prime>T_{\text{risk}},N_0, y_0\right.\right)$ is the conditional truncated Beta distribution:
\begin{align}
    \mathbb{P}\left(r^\prime\left| r^\prime>T_{\text{risk}},N_0, y_0\right.\right) & =\dfrac{\mathbb{P}\left(r^\prime\left| N_0, y_0\right.\right)}{\int_{T_{\text{risk}}}^{1}d\tilde{r}\mathbb{P}\left(\tilde{r}\left| N_0, y_0\right.\right)}.
\end{align}
This method aims to choose the smallest $N_{1}$ so that, if $r>T_{\text{risk}}$, then taking $N_{1}$ samples will result in an orange or red posterior with $95\%$ probability.

Our implementation in R uses a log transformation before calculating certain parts (Appendix~\ref{app:technical}) to improve numerical precision. We use the normal approximation throughout because it shortens calculation run time and also gives smooth results (see Appendix~\ref{app:Normal-approximation}). Note that the normal approximation gives conservative results, meaning that the sample sizes are sufficient, but may be slightly higher than necessary.

Our recommended sampling is an weighted average. The sampling volume to detect a certain fixed rate above $T_{\text{risk}}$ with 95\% probability would deviate from the calculated `minimum sampling'. For rates very close and just above $T_{\text{risk}}$, the sampling required will increase compared to the weighted average. Meanwhile, for rates far above $T_{\text{risk}}$, the sampling required will decrease.

Note that the changing parcel volumes do not impact the risk estimation. The method require certain minimum numbers of samples for statistical significance to characterise the leakage rate distribution in the pathways, determined by the sample size calculation. This leakage rate is insensitive to changing parcel volumes --- i.e. we assume that the percentage is the same even as parcel volumes fluctuate. 

\subsubsection{Step 3: Conduct sampling over the reporting period}

Given the assumptions of our model, the sampling should be random. After sampling $N_{1}$ items, we find some $y_{1}$ contaminated items.

\subsubsection{Step 4: Update our belief regarding the leakage rate\label{subsec:Updating-our-belief}}

As we acquire more data, we need to adjust our corresponding belief by updating the beta distribution. After sampling $N_{1}$ items and finding $y_{1}$ contaminated items, the updated belief distribution is given by the beta distribution $r\left|N_0, y_0, N_1, y_1\right.\sim\text{Beta}(\alpha_{1},\beta_{1})$ as given in Eq.~\eqref{eq:posterior_r}, i.e.  with updated posterior parameters \cite[\S2.4]{gelmanbda04}:
\begin{align}
\alpha_{1} & =y_{1}+y_0 +0.5\label{eq:updatedalpha}\\
\beta_{1} & =N_{1}-y_{1}+N_0-y_0 + 0.5.\label{eq:updatedbeta}
\end{align}

\subsubsection{Step 5: Evaluate the current status of the pathway}

Given the updated belief distribution $\text{Beta}(\alpha_{1},\beta_{1})$, we can now assign an updated status, based on the traffic light system in Figure~\ref{fig:colour_statuses} and Eq.~\eqref{eq:colour_statues}. That is: 
\begin{itemize}
\item The updated status is green if $\mathbb{P}\left(r<T_{\text{change}}\left(N_0, y_0\right)\left|N_0, y_0, N_1, y_1\right.\right)\ge0.95$.
\item The updated status is orange if $\mathbb{P}\left(r<T_{\text{risk}}\left|N_0, y_0, N_1, y_1\right.\right)\ge0.95$.
\item The updated status is red otherwise.
\end{itemize}
(Note that $T_{\text{change}}$ is defined relative to the prior data.)

If the updated status is red, then the pathway is no longer confidently low risk and should be dealt with in some manner. This could mean separating out the high-risk subpathway or applying targeted investigation.

\subsubsection{Step 6: Repeat}

To continue the process, we append the newly collected data to our prior data, and remove appropriately old data. If the pathway now appears to have a red status, our method will not work, and the high-risk subpathway needs to be removed before continuing back to step 1.

\section{Simulated scenarios}\label{sec:Scenario-examples}

To validate our sampling procedure, we ran simulations of the possible outcomes. We considered a scenario where the leakage rate remains acceptable (``routine''), a scenario where the leakage rate rises above the risk threshold $T_{\text{risk}}$ (``risky''), and a scenario where the leakage rate is very low below the risk threshold (``very low risk''). We compared our method with two other methods, that of \citeA{Robinson2011}, and fixed sampling at 600 inspections \cite{ISPM_31,Whyte2009_ISPM}.

\subsection{Simulation process}

We simulate the pathway by using fixed true rates $r_{\text{true}}$ for the approaching contamination. For a given number of inspections $N_{1}$ (chosen by the methods), we run $N_{1}$ random Bernoulli trials with rate $r_{\text{true}}$, leading to a list of $N_{1}$ inspection results, with some $y_{1}$ detected contamination. The process is repeated. Some model-specific parameters are given in Table~\ref{tab:Model_parameters}. Note that we call a single reporting period a ``quarter'' (i.e., three months), but this is simply for illustration.

\begin{table*}[h]
\centering  \caption{Model-specific parameters used as part of the simulation process. *Note: While \protect\citeA{Robinson2011}'s method technically requires a pathway size for calculations, the sample size is independent of the pathway size for sufficiently large pathways and if we are not in the 100\%-inspection stage. **The sampling size determination of \protect\citeA{ISPM_31,Whyte2009_ISPM} depends on detection level, confidence level, and efficacy. Given those parameters (0.5\%, 95\% and 100\% respectively), the sampling volume is fixed (technically to 598) and does not vary in relation to past sample results.}
\label{tab:Model_parameters}
\centering{}%
\begin{tabular}{cccc}
\toprule
\textbf{Parameter} & \textbf{Our method} & \textbf{\citeA{Robinson2011}'s method }& \textbf{600 samples method}\\
\toprule
$T_{\text{risk}}$ & $0.5\%$ & $0.5\%$ (risk-cutoff) & $0.5\%$ (detection level) \\
$T_{\text{change}}$ & one-sided $95\%$ credible interval & N/A & N/A \\
$N_{0}$ (prior) & $10000$ & $10000$ & N/A \\
$y_{0}$ (prior) & $6$ & $6$ & N/A \\
Pathway size & N/A & N/A* & N/A \\
Sample size & Variable, calculated & Variable, calculated & 600, fixed** \\
\bottomrule
\end{tabular}
\end{table*}

\subsubsection{Thresholds}

As the 600-sample method is based on being able to detect contamination rates above $0.5\%$ with $95\%$ probability \cite{ISPM_31,Whyte2009_ISPM}, we take $0.5\%$ as the low-risk threshold $T_{\text{risk}}$, which is also known as the risk-cutoff in \citeA{Robinson2011}. Meanwhile, we take $T_{\text{change}}$ to be the $95\%$ one-sided credible interval (c.f. Eq.~\eqref{eq:T_change_definition}). 

\subsubsection{Prior data}

Our method and \citeA{Robinson2011}'s method both require some prior data. We assume that during the last two reporting periods, there were 5000 samples each, with $3$ detected leakages each, for a total of $N_{0}=10000$ samples and $y_{0}=6$ detected non-compliance. 

These two methods also require prior data for each subsequent iteration. We chose to keep the most recent data from the last two reporting periods, while discarding older samples.

\subsubsection{Pathway size}

As part of \citeA{Robinson2011}'s method, there is regime where 100\% of samples are inspected, which would necessitate us having some full pathway size. However, for simulations, we end the simulations before this happens, so this is not necessary.

\subsection{Routine scenario\label{subsec:Routine-examples}}

\begin{figure*}[h]
\centerline{\includegraphics[width=\textwidth]{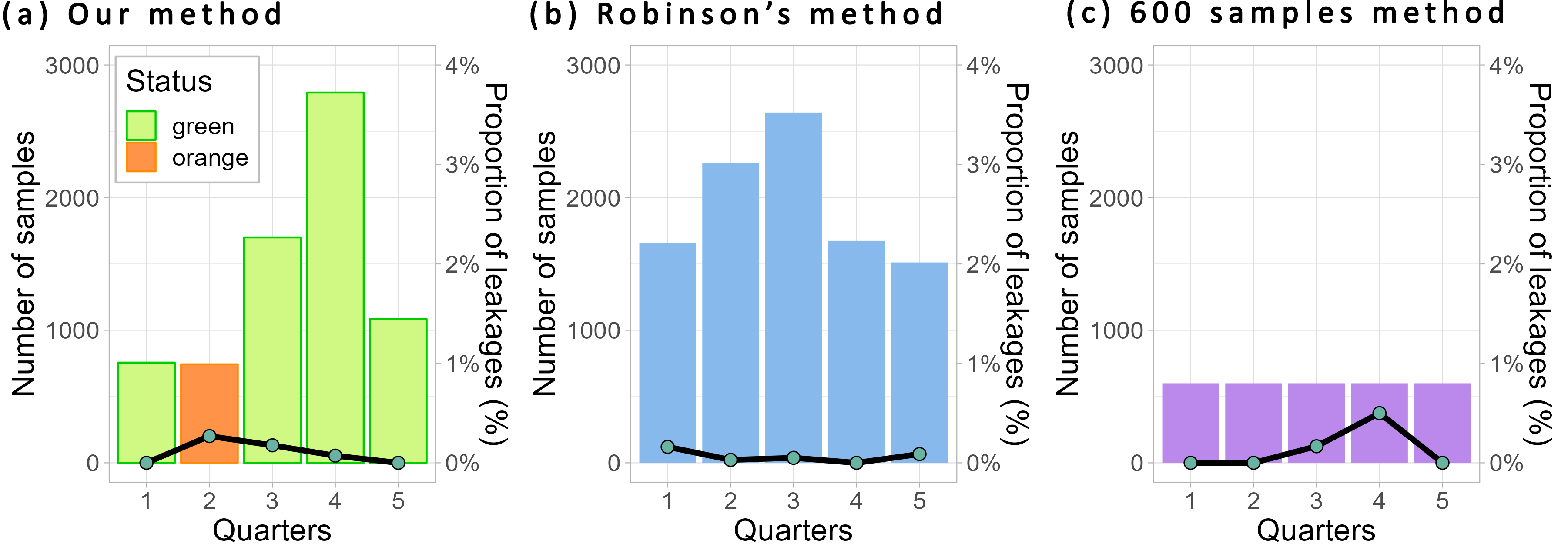}}
\caption{An illustrative example of routine sampling (with the underlying leakage rate fixed at $0.12\%$). (a) Our method; (b) \protect\citeA{Robinson2011}'s method; (c) Fixed 600 samples method. 
The bars' height shows the amount of sampling done in that quarter (see left axes). The black line shows the proportion of leakages per quarter (see right axes).
\label{fig:scenario_routine}}
\end{figure*}

First, consider the scenario where the contamination rate $r_{\text{true}}=0.12\%$ is fixed below $T_{\text{risk}}=0.5\%$ for the duration of the study: the results are shown in Figure~\ref{fig:scenario_routine}.
Notably, \citeA{Robinson2011}'s method requires the most sampling, our method the second most, while the 600-sampling method results in the least sampling.

If we focus on our method (Fig.~\ref{fig:scenario_routine}(a)), we see that the number of samples requested during each quarter (reporting period) can vary widely. The sampling process is stochastic, so occasionally more leakages will be detected, leading to an orange belief (even though the true rate is fixed in these simulations). As a result, more samples are recommended for the subsequent quarters (e.g. quarters 3 and 4). The resulting green statuses allows the recommended sampling to reduce again.

It can be seen that the benefit of more sampling, as in \citeA{Robinson2011}'s method, is that the proportion of detected leakages during each quarter is small and more stable (Fig.~\ref{fig:scenario_routine}(b)). In contrast, with only a 600 sample volume, finding a few extra leakages causes a big spike in apparent leakage rate (quarter 4 in Fig.~\ref{fig:scenario_routine}(c)), which could lead one to erroneously believe that the leakage rate has increased about the low-risk threshold $T_\text{risk}$.

\subsection{Risky scenario\label{subsec:Risky-examples}}

\begin{figure*}[h]
\begin{centering}
\includegraphics[width=1\textwidth]{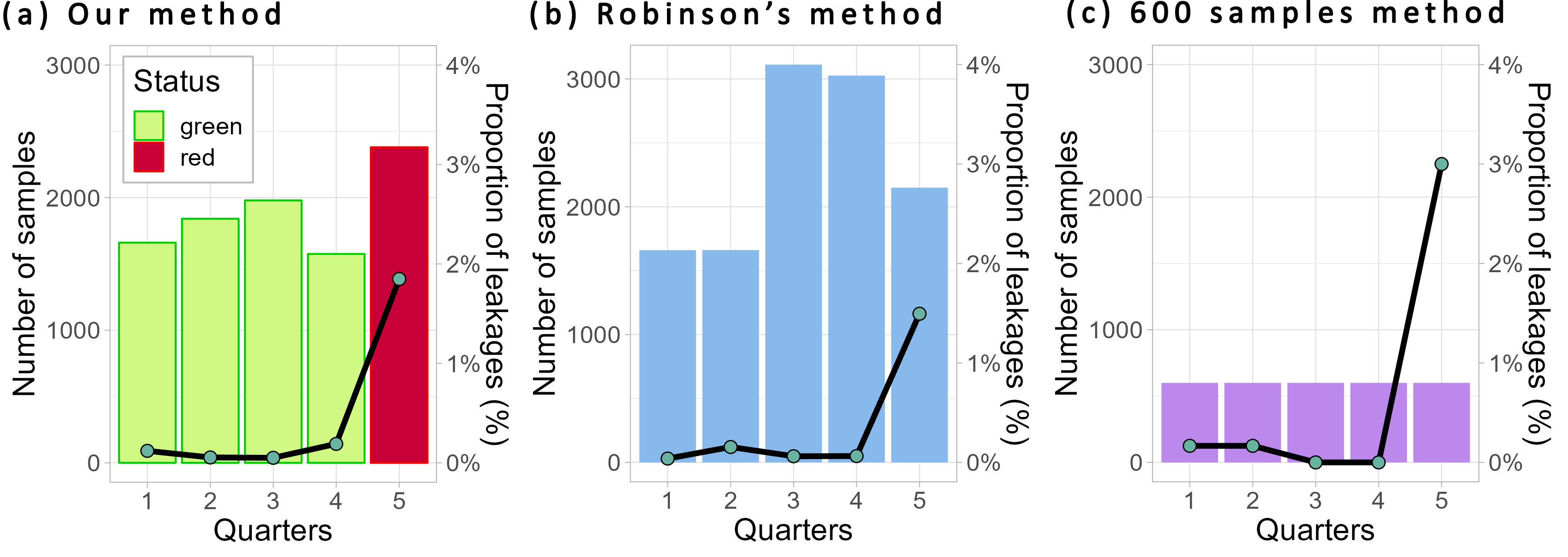}
\par\end{centering}
\caption{An illustrative example of routine sampling (underlying leakage rate is fixed at $0.12\%$) followed by rise in contamination (to $2\%$) in quarter 5. (a) Our method; (b) \protect\citeA{Robinson2011}'s method; (c) Fixed 600 samples method. The bars' height shows the amount of sampling done in that quarter (see left axis). The black line shows the proportion of leakages per quarter (see right axis).
\label{fig:scenario_risky}}
\end{figure*}

In an alternative scenario, suppose if the underlying leakage rate increases from quarter $5$ onwards, and remains elevated at $2\%$. The example of the result surveys is shown in Figure~\ref{fig:scenario_risky}. We see that in all methods detect a sharp rise in contaminated items.

\subsection{Very low risk scenario}

\begin{figure*}[h]
\begin{centering}
\includegraphics[width=1\textwidth]{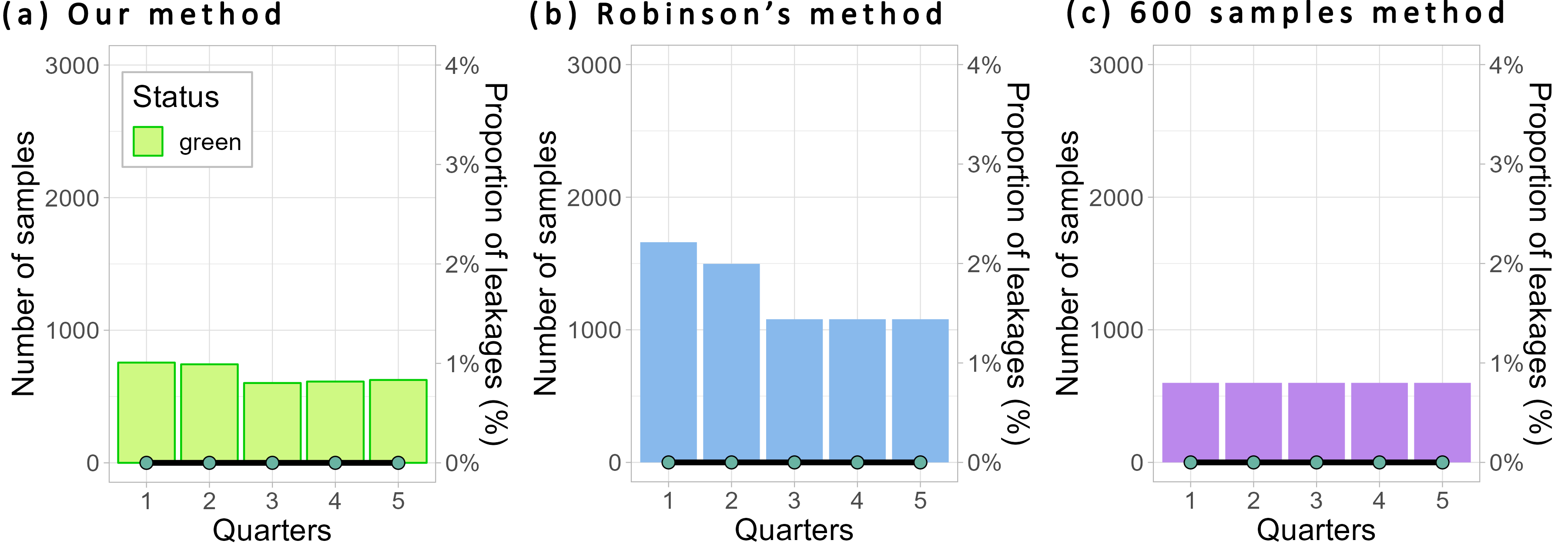}
\par\end{centering}
\caption{An illustrative example of routine sampling when the risk is very low (the underlying leakage rate is fixed at $0.01\%$). (a) Our method; (b) \protect\citeA{Robinson2011}'s method; (c) Fixed 600 samples method. The bars' height shows the amount of sampling done in that quarter (see left axes). The black line shows the proportion of leakages per quarter (see right axes).
\label{fig:scenario_very_low_risk} }
\end{figure*}

See Figure~\ref{fig:scenario_very_low_risk}, where we consider if the risk is very quite low. In this scenario, our model recommends sampling volumes close to 600, roughly aligning with the 600-sampling method. (Alternatively, this can allow us to interpret a 600-sampling volume as being suitable for a situation when the expected risk in the pathway is many orders of magnitude smaller than the unacceptable risk threshold.)

Meanwhile, \cite{Robinson2011}'s method recommends almost double the number of samples. 

\subsection{Probability of the recommended minimum sampling value to detect different rises in contamination rate}

The scenario examples in Figures~\ref{fig:scenario_routine}, \ref{fig:scenario_risky}, and \ref{fig:scenario_very_low_risk} were just that --- examples. Due to the randomness inherent in the sampling, the actual results would differ across each iteration.

So here, we calculate the proportion of colour statuses that result from inspection, given different contamination rates, over 100 iterations. We start with the prior $N_{0}=10000$, $y_{0}=6$, with $T_{\text{risk}}=0.5\%$ and $T_{\text{change}}$ set at the $95\%$ one-sided credible interval. If so, the recommended minimum sampling from our model for the next time period is $N_{1}=756$. We consider true rates from $0.0\%$ to $5\%$.

\begin{figure*}
\begin{centering}
\includegraphics[width=0.9\textwidth]{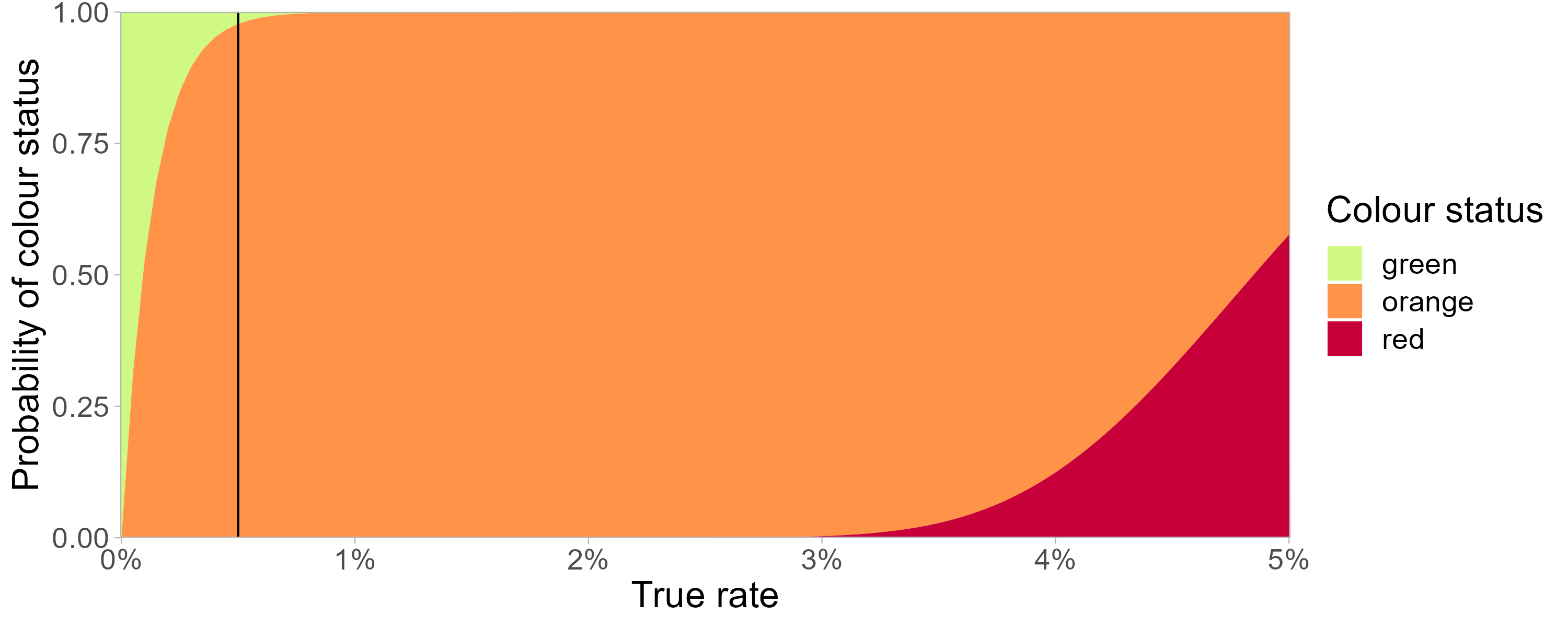}
\par\end{centering}
\caption{Posterior colour status after sampling the recommended minimum sample size, given different contamination rates ($0.0\%$ to $5\%$), in one reporting period. The black vertical line marks the low-risk threshold $T_\text{risk}$ at $0.5\%$.\label{fig:probability_detection}}
\end{figure*}

In Figure~\ref{fig:probability_detection}, we see that the status is more frequently green at very low risk value, before becoming primarily orange by the time the low-risk threshold is passed. Frequent red status only starts occurring after the leakage rate rises above $3\%$. 

\section{Discussion}\label{sec:Discussion}

\subsection{Summary of findings}

In this paper, we introduced a methodology to calculate minimal sample sizes required to be confident that a pathway is low risk, i.e.  that the rate of non-compliance is below the risk threshold.

Using our methodology, the minimum number of samples for each sampling period depends on the past rate of non-compliance and on the risk threshold. In general, higher risk thresholds and lower past rates of non-compliance each lead to lower sampling in the future. We provide some simulations to demonstrate how survey effort could be used and how our method behaves when the rate of non-compliance in a pathway changes. 

The advantage of our new method is that we can reduce sampling effort compared to the method by \citeA{Robinson2011}, and is adaptive to changes in risk unlike the fixed sampling number, while maintaining confidence in our risk. 

\subsection{Comparison with Robinson et al.'s method}

Our method has many similarities with that of \citeA{Robinson2011}. We both use a beta distribution to estimate leakage rates, and the update method is the same (following the work of \citeA{cai2005}). However, the approach previously taken to calculate recommended future sampling is different. Statistically speaking, their analysis is premised on the hypothesis that the non-compliance rate is above the given risk threshold. Then, the number of samples is formulated to \emph{reject} this assumption in the case that it is false.

In contrast, we took a different approach. Rather, the null hypothesis should be that the leakage rate remains constant --- the number of samples is chosen to reject this whenever the rate goes above the specified risk level. This allows us to reduce the amount of sampling in a low risk low-search setting, while maintaining statistical confidence about the underlying leakage risk.

As such, our method requires less sampling effort when the risk is low when compared to the method by \citeA{Robinson2011}. Because we have less sampling, we also incorporated an aspect of change detection to give confidence that risks are below certain thresholds.

\subsection{Comparison with fixed 600 samples}

The work by the ISPM \cite{ISPM_31,Whyte2009_ISPM} sets out a range of possible recommended sampling volumes, given different consignment sizes, confidence levels, tolerance levels etc. The 600 sample size is such that there is a 95\% probability of detecting contamination if the infestation rate is 0.5\% or above. Technically, this is designed for per-consignment inspection; here, we applied it to the entire low-risk pathway, using the same premise. Notably, this method does not use any past data.

When the leakage rate is very low, our method would suggest sample sizes close to 600. However, in most other cases, our method suggested larger sample sizes. The disadvantage of having a sample size is the lowered statistical power.

\subsection{Model assumptions and limitations}

Our method makes some assumptions. Notably: the sampling should be chosen randomly. We assume that the contamination/non-compliance is also distributed randomly across the pathway. We also assume that inspection sensitivity is 100\%. We also assume that the low-risk pathway is large, so that we use the binomial distribution to represent the leakages, rather than a hyper-geometric distribution. We also assume that we can vary sampling volumes across different reporting periods. We do not stratify the low-risk pathway (e.g., by importer \cite{Powell2015}).

\subsection{Choosing model parameters}

Our method requires several parameters that must be chosen, which are described in this section.

\subsubsection{Choosing the thresholds}

Firstly, the low-risk threshold $T_{\text{risk}}$ is not defined by our method. Instead, it should be chosen by regulators and managers, based on what they consider to be no longer acceptably low-risk. 

In practice, we can vary the definition of $T_{\text{change}}$, such that it is either more or less sensitive to changes in the rate. This in turn affects the amount of sampling the method recommends: if $T_{\text{change}}$ is set to the one-sided 96\% credible interval (instead of 95\%), then more sampling is recommended. This is illustrated in the right plot of Figure~\ref{fig:opt_samples_plot}.

\subsubsection{Choosing prior data sizes\label{subsec:Choosing-prior-data}}

The prior sample size is different from the future sample size. At the start of each time period, we must choose a certain amount of prior samples to keep. For example, we could choose a fixed number of prior samples (e.g. the most recent 5000 samples); alternatively we could choose the prior samples within a certain time period (say, the last 6 months).

There is a minimum condition: the prior data kept must be such that it is classed as low-risk, i.e. the $95\%$ credible interval of our leakage rate distribution is below $T_{\text{risk}}$.

Beyond this minimum necessity, the precise choice of prior data size depends on a number of factors. The most important factor to consider is how confident we are about the change of the rate in the pathways. If we have reason to believe that the leakage rates are stable across time, then more prior samples, or prior data going further back in time, can be included to reflect this.

\begin{figure*}
\begin{centering}
\includegraphics[width=1\textwidth]{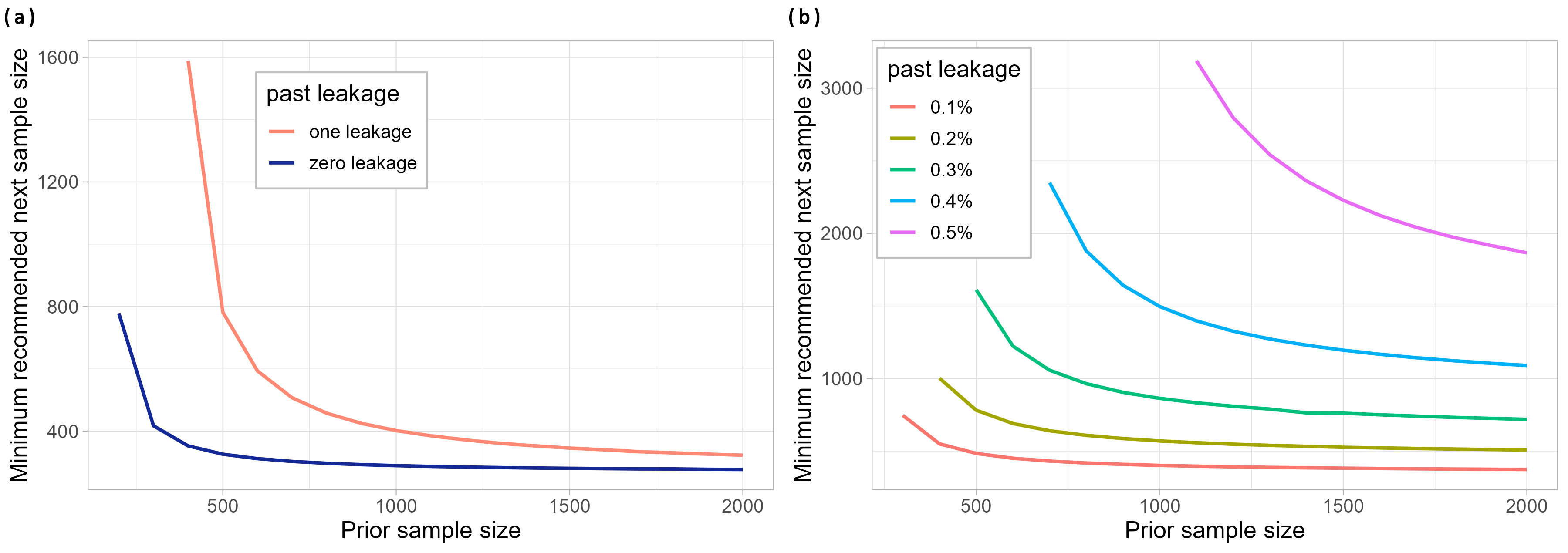}
\par\end{centering}
\caption{The minimum recommended number of samples different prior sample sizes with $T_{\text{risk}}=1\%$ and $T_{\text{change}}$ set at 95\% credible interval of the prior distribution. (a) With either zero leakages or one leakage, (b) with fixed past leakage rates.\label{fig:optimal_samples_given_prior_N}}
\end{figure*}

Aside from this, another factor to consider is how past prior sample size affects the future sample size. For example, in Figure~\ref{fig:optimal_samples_given_prior_N}, we see that the smallest prior sample sizes will lead to the largest future sample sizes (largest recommended minimum next sample size). 

Initially increasing the number of prior samples leads to a reduction in future sample sizes. At a certain point, this trend reverses and
increasing prior sample sizes subsequently means larger future sample sizes are needed. More generally, Figure~\ref{fig:optimal_samples_given_prior_N} shows that it is typically not ideal to have the absolute minimum number of prior samples possible, whilst there is more flexibility at larger prior sample sizes as they do not change the future sample size as much.

\subsubsection{If there was no prior data}

If there is no prior data, then we must first estimate the prior leakage rate through various methods. If this pathway is very different from other pathways for which data are available, then expert elicitation can be used to create the beta distribution encoding the experts' belief about the leakage rate.

If this pathway is similar to other pathways (for various external reasons, or from expert opinion), then we could use historical data from other pathways and proceed from there. As the reporting periods continue, the old historical data from other pathways will eventually be discarded.

\subsubsection{When the posterior or prior status becomes red}

Our method assumes that the initial belief is low risk (green). Therefore, the method to generate the minimum recommended sample volumes for the next time period does not apply if the posterior or prior status becomes red. This is because the survey design calculates the minimum sample size to detect when the rate goes above the $T_{\text{risk}}$ value; if the prior or posterior status is red, then technically zero samples are needed in the next time period as red-status has already been obtained.

If the status of the posterior or prior beta distribution of the leakage rate becomes red --- and thus potentially no longer low risk --- and we wish to continue using the survey design method, then the subpathway where the leakages are detected should be removed from the rest of the pathway. That is, in general, the increased risk should be within a particular subpathway (e.g. items from a certain country, of certain type, etc.), rather than the full pathway. To continue the survey method, the excess leakages and samples from this subpathway should be removed from the rest of the data. The pathway with increased risk should be analysed separately as it no longer classed as low risk and outside the purview of the baseline survey.

\section{Conclusion}\label{sec:Conclusion}

We developed a method that uses prior data to inform future survey volumes on low-risk pathways. Our methods works by focusing on determining whether the risk is below a threshold, rather than estimating the risk precisely, which allows us to use relatively lower sampling volumes. If we start to find more contamination, then the sampling volume increases to confirm whether or not this rise in risk is true. This method could help regulators efficiently and effectively allocate inspection effort.

\bmsection*{Author contributions}

All authors developed the method, with A.R. and H.B. providing mathematical support. T.P.L., T.K.W., and C.M.B. wrote the R code, conducted the simulations, produced the figures, and wrote various parts of the original manuscript. All authors provided manuscript editorial reviews.

\bmsection*{Acknowledgements}
We acknowledge the input and advice provided by Stephen Larcombe, Tina Safaris, Tonci Padovan, Nin Hyne, Adil Waqas and Andrew Brettargh. 

\bmsection*{Financial disclosure}

We acknowledge the financial and other support provided by the Australian Department of Agriculture, Fisheries and Forestry (previously known as the Australian Department of Agriculture, Water and Environment),  the New Zealand Ministry for Primary Industries, the University of Melbourne, and the Australian Research Council. 

\bmsection*{Conflict of interest}

The authors declare no potential conflict of interests.

\bibliography{bibliography}

\begin{thebibliography}{}

\bibitem [\protect \citeauthoryear {%
{Australian Department of Agriculture, Water and the Environment}%
}{%
{Australian Department of Agriculture, Water and the Environment}%
}{%
{\protect \APACyear {2021}}%
}]{%
AusHorticulture}
\APACinsertmetastar {%
AusHorticulture}%
\begin{APACrefauthors}%
{Australian Department of Agriculture, Water and the Environment}.%
\end{APACrefauthors}%
\unskip\
\newblock
\APACrefYearMonthDay{2021}{}{}.
\newblock
\APACrefbtitle {Inspection of horticulture for export.} {Inspection of horticulture for export.}
\newblock
\begin{APACrefURL} \url{https://www.agriculture.gov.au/sites/default/files/sitecollectiondocuments/biosecurity/export/plants-plant-products/plant-exports-manual/guideline-inspection-horticulture-export.pdf} \end{APACrefURL}
\PrintBackRefs{\CurrentBib}

\bibitem [\protect \citeauthoryear {%
Burne%
}{%
Burne%
}{%
{\protect \APACyear {2019}}%
}]{%
BMSB_NZ}
\APACinsertmetastar {%
BMSB_NZ}%
\begin{APACrefauthors}%
Burne, A\BPBI R.%
\end{APACrefauthors}%
\unskip\
\newblock
\APACrefYearMonthDay{2019}{{\APACmonth{06}}}{}.
\newblock
\APACrefbtitle {Pest risk assessment: Halyomorpha halys (Brown marmorated stink bug)} {Pest risk assessment: Halyomorpha halys (brown marmorated stink bug)}\ \APACbVolEdTR{}{}.
\newblock
\APACaddressInstitution{}{Ministry for Primary Industries, New Zealand}.
\newblock
\begin{APACrefURL} \url{https://www.mpi.govt.nz/dmsdocument/38075-Pest-risk-assessment-Halyomorpha-halys-Brown-marmorated-stink-bug-Technical-Paper} \end{APACrefURL}
\newblock
\APACrefnote{Biosecurity New Zealand Technical Paper No:2019/43}
\PrintBackRefs{\CurrentBib}

\bibitem [\protect \citeauthoryear {%
{Department of Agriculture, Fisheries and Forestry}%
}{%
{Department of Agriculture, Fisheries and Forestry}%
}{%
{\protect \APACyear {2023}}%
}]{%
CAL_containers}
\APACinsertmetastar {%
CAL_containers}%
\begin{APACrefauthors}%
{Department of Agriculture, Fisheries and Forestry}.%
\end{APACrefauthors}%
\unskip\
\newblock
\APACrefYearMonthDay{2023}{}{}.
\newblock
\APACrefbtitle {{Country Action List (CAL)}.} {{Country Action List (CAL)}.}
\newblock
\begin{APACrefURL} \url{https://www.agriculture.gov.au/biosecurity-trade/import/arrival/pests/cal} \end{APACrefURL}
\newblock
\APACrefnote{Accessed 2023}
\PrintBackRefs{\CurrentBib}

\bibitem [\protect \citeauthoryear {%
Diagne%
\ \protect \BOthers {.}}{%
Diagne%
\ \protect \BOthers {.}}{%
{\protect \APACyear {2021}}%
}]{%
Diagne2021}
\APACinsertmetastar {%
Diagne2021}%
\begin{APACrefauthors}%
Diagne, C.%
, Leroy, B.%
, Vaissi{\`{e}}re, A\BHBI C.%
, Gozlan, R\BPBI E.%
, Roiz, D.%
, Jari{\'{c}}, I.%
\BDBL {}Courchamp, F.%
\end{APACrefauthors}%
\unskip\
\newblock
\APACrefYearMonthDay{2021}{mar}{}.
\newblock
{\BBOQ}\APACrefatitle {High and rising economic costs of biological invasions worldwide} {High and rising economic costs of biological invasions worldwide}.{\BBCQ}
\newblock
\APACjournalVolNumPages{Nature}{592}{7855}{571--576}.
\newblock
\begin{APACrefDOI} 10.1038/s41586-021-03405-6 \end{APACrefDOI}
\PrintBackRefs{\CurrentBib}

\bibitem [\protect \citeauthoryear {%
Fantle-Lepczyk%
\ \protect \BOthers {.}}{%
Fantle-Lepczyk%
\ \protect \BOthers {.}}{%
{\protect \APACyear {2022}}%
}]{%
FantleLepczyk2022}
\APACinsertmetastar {%
FantleLepczyk2022}%
\begin{APACrefauthors}%
Fantle-Lepczyk, J\BPBI E.%
, Haubrock, P\BPBI J.%
, Kramer, A\BPBI M.%
, Cuthbert, R\BPBI N.%
, Turbelin, A\BPBI J.%
, Crystal-Ornelas, R.%
\BDBL {}Courchamp, F.%
\end{APACrefauthors}%
\unskip\
\newblock
\APACrefYearMonthDay{2022}{feb}{}.
\newblock
{\BBOQ}\APACrefatitle {Economic costs of biological invasions in the United States} {Economic costs of biological invasions in the united states}.{\BBCQ}
\newblock
\APACjournalVolNumPages{Science of The Total Environment}{806}{}{151318}.
\newblock
\begin{APACrefDOI} 10.1016/j.scitotenv.2021.151318 \end{APACrefDOI}
\PrintBackRefs{\CurrentBib}

\bibitem [\protect \citeauthoryear {%
Fenn-Moltu%
\ \protect \BOthers {.}}{%
Fenn-Moltu%
\ \protect \BOthers {.}}{%
{\protect \APACyear {2022}}%
}]{%
FennMoltu2022}
\APACinsertmetastar {%
FennMoltu2022}%
\begin{APACrefauthors}%
Fenn-Moltu, G.%
, Ollier, S.%
, Caton, B.%
, Liebhold, A\BPBI M.%
, Nahrung, H.%
, Pureswaran, D\BPBI S.%
\BDBL {}Bertelsmeier, C.%
\end{APACrefauthors}%
\unskip\
\newblock
\APACrefYearMonthDay{2022}{nov}{}.
\newblock
{\BBOQ}\APACrefatitle {Alien insect dispersal mediated by the global movement of commodities} {Alien insect dispersal mediated by the global movement of commodities}.{\BBCQ}
\newblock
\APACjournalVolNumPages{Ecological Applications}{33}{1}{}.
\newblock
\begin{APACrefDOI} 10.1002/eap.2721 \end{APACrefDOI}
\PrintBackRefs{\CurrentBib}

\bibitem [\protect \citeauthoryear {%
Gelman%
\ \protect \BOthers {.}}{%
Gelman%
\ \protect \BOthers {.}}{%
{\protect \APACyear {2021}}%
}]{%
gelmanbda04}
\APACinsertmetastar {%
gelmanbda04}%
\begin{APACrefauthors}%
Gelman, A.%
, Carlin, J\BPBI B.%
, Stern, H\BPBI S.%
, Dunson, D\BPBI B.%
, Vehtari, A.%
\BCBL {}\ \BBA {} Rubin, D\BPBI B.%
\end{APACrefauthors}%
\unskip\
\newblock
\APACrefYear{2021}.
\newblock
\APACrefbtitle {Bayesian Data Analysis} {Bayesian data analysis}\ (\PrintOrdinal{3rd}\ \BEd).
\newblock
\APACaddressPublisher{}{Chapman and Hall}.
\PrintBackRefs{\CurrentBib}

\bibitem [\protect \citeauthoryear {%
Herrick%
}{%
Herrick%
}{%
{\protect \APACyear {2011}}%
}]{%
Herrick2011}
\APACinsertmetastar {%
Herrick2011}%
\begin{APACrefauthors}%
Herrick, C.%
\end{APACrefauthors}%
\unskip\
\newblock
\APACrefYearMonthDay{2011}{}{}.
\newblock
\APACrefbtitle {Brown Marmorated Stink Bug Causes \$37 Million In Losses To Mid-Atlantic Apple Growers.} {Brown marmorated stink bug causes \$37 million in losses to mid-atlantic apple growers.}
\newblock
\begin{APACrefURL} \url{https://www.growingproduce.com/fruits/apples-pears/brown-marmorated-stink-bug-causes-37-million-in-losses-to-mid-atlantic-apple-growers/} \end{APACrefURL}
\newblock
\APACrefnote{Accessed 2023}
\PrintBackRefs{\CurrentBib}

\bibitem [\protect \citeauthoryear {%
Hester%
\ \protect \BOthers {.}}{%
Hester%
\ \protect \BOthers {.}}{%
{\protect \APACyear {2020}}%
}]{%
Hester2020}
\APACinsertmetastar {%
Hester2020}%
\begin{APACrefauthors}%
Hester, S.%
, Rossiter, A.%
, Robinson, A.%
, Sibley, J.%
, Woolcott, B.%
, Aston, C.%
\BCBL {}\ \BBA {} Hanea, A.%
\end{APACrefauthors}%
\unskip\
\newblock
\APACrefYearMonthDay{2020}{}{}.
\newblock
\APACrefbtitle {{CBIS/CSP} sensitivity: incorporating pre-border information analysis} {{CBIS/CSP} sensitivity: incorporating pre-border information analysis}\ \APACbVolEdTR{}{}.
\newblock
\APACaddressInstitution{}{CEBRA}.
\newblock
\begin{APACrefURL} \url{https://cebra.unimelb.edu.au/__data/assets/pdf_file/0018/3811023/170608-Final-Report-for-web.pdf} \end{APACrefURL}
\PrintBackRefs{\CurrentBib}

\bibitem [\protect \citeauthoryear {%
Hudgins%
, Koch%
, Ambrose%
\BCBL {}\ \BBA {} Leung%
}{%
Hudgins%
\ \protect \BOthers {.}}{%
{\protect \APACyear {2022}}%
}]{%
Hudgins2022}
\APACinsertmetastar {%
Hudgins2022}%
\begin{APACrefauthors}%
Hudgins, E\BPBI J.%
, Koch, F\BPBI H.%
, Ambrose, M\BPBI J.%
\BCBL {}\ \BBA {} Leung, B.%
\end{APACrefauthors}%
\unskip\
\newblock
\APACrefYearMonthDay{2022}{}{}.
\newblock
{\BBOQ}\APACrefatitle {Hotspots of pest-induced {US} urban tree death, 2020–2050} {Hotspots of pest-induced {US} urban tree death, 2020–2050}.{\BBCQ}
\newblock
\APACjournalVolNumPages{Journal of Applied Ecology}{59}{5}{1302-1312}.
\newblock
\begin{APACrefDOI} 10.1111/1365-2664.14141 \end{APACrefDOI}
\PrintBackRefs{\CurrentBib}

\bibitem [\protect \citeauthoryear {%
{International Standards for Phytosanitary Measures}%
}{%
{International Standards for Phytosanitary Measures}%
}{%
{\protect \APACyear {2008}}%
}]{%
ISPM_31}
\APACinsertmetastar {%
ISPM_31}%
\begin{APACrefauthors}%
{International Standards for Phytosanitary Measures}.%
\end{APACrefauthors}%
\unskip\
\newblock
\APACrefYearMonthDay{2008}{}{}.
\newblock
\APACrefbtitle {{ISPM No. 31 Methodologies for Sampling of Consignments}} {{ISPM No. 31 Methodologies for Sampling of Consignments}}\ \APACbVolEdTR{}{}.
\newblock
\APACaddressInstitution{}{IPPC}.
\newblock
\begin{APACrefURL} \url{https://www.ippc.int/en/publications/83473/} \end{APACrefURL}
\PrintBackRefs{\CurrentBib}

\bibitem [\protect \citeauthoryear {%
Kriticos%
\ \protect \BOthers {.}}{%
Kriticos%
\ \protect \BOthers {.}}{%
{\protect \APACyear {2017}}%
}]{%
Kriticos2017}
\APACinsertmetastar {%
Kriticos2017}%
\begin{APACrefauthors}%
Kriticos, D\BPBI J.%
, Kean, J\BPBI M.%
, Phillips, C\BPBI B.%
, Senay, S\BPBI D.%
, Acosta, H.%
\BCBL {}\ \BBA {} Haye, T.%
\end{APACrefauthors}%
\unskip\
\newblock
\APACrefYearMonthDay{2017}{may}{}.
\newblock
{\BBOQ}\APACrefatitle {The potential global distribution of the brown marmorated stink bug, Halyomorpha halys, a critical threat to plant biosecurity} {The potential global distribution of the brown marmorated stink bug, halyomorpha halys, a critical threat to plant biosecurity}.{\BBCQ}
\newblock
\APACjournalVolNumPages{Journal of Pest Science}{90}{4}{1033--1043}.
\newblock
\begin{APACrefDOI} 10.1007/s10340-017-0869-5 \end{APACrefDOI}
\PrintBackRefs{\CurrentBib}

\bibitem [\protect \citeauthoryear {%
T.~Leskey%
\ \BBA {} Hamilton%
}{%
T.~Leskey%
\ \BBA {} Hamilton%
}{%
{\protect \APACyear {2010}}%
}]{%
Leskey2010}
\APACinsertmetastar {%
Leskey2010}%
\begin{APACrefauthors}%
Leskey, T.%
\BCBT {}\ \BBA {} Hamilton, G.%
\end{APACrefauthors}%
\unskip\
\newblock
\APACrefYearMonthDay{2010}{}{}.
\newblock
\APACrefbtitle {Brown marmorated stink bug working group meeting.} {Brown marmorated stink bug working group meeting.}
\newblock
\begin{APACrefURL} \url{http://projects.ipmcenters.org/Northeastern/FundedProjects/ReportFiles/Pship2010/Pship2010-Leskey-ProgressReport-237195-Meeting-2010_11_17.pdf} \end{APACrefURL}
\newblock
\APACrefnote{Accessed 2023}
\PrintBackRefs{\CurrentBib}

\bibitem [\protect \citeauthoryear {%
T\BPBI C.~Leskey%
\ \BBA {} Nielsen%
}{%
T\BPBI C.~Leskey%
\ \BBA {} Nielsen%
}{%
{\protect \APACyear {2018}}%
}]{%
Leskey2018}
\APACinsertmetastar {%
Leskey2018}%
\begin{APACrefauthors}%
Leskey, T\BPBI C.%
\BCBT {}\ \BBA {} Nielsen, A\BPBI L.%
\end{APACrefauthors}%
\unskip\
\newblock
\APACrefYearMonthDay{2018}{jan}{}.
\newblock
{\BBOQ}\APACrefatitle {Impact of the Invasive Brown Marmorated Stink Bug in North America and Europe: History, Biology, Ecology, and Management} {Impact of the invasive brown marmorated stink bug in north america and europe: History, biology, ecology, and management}.{\BBCQ}
\newblock
\APACjournalVolNumPages{Annual Review of Entomology}{63}{1}{599--618}.
\newblock
\begin{APACrefDOI} 10.1146/annurev-ento-020117-043226 \end{APACrefDOI}
\PrintBackRefs{\CurrentBib}

\bibitem [\protect \citeauthoryear {%
{Ministry for Primary Industries}%
}{%
{Ministry for Primary Industries}%
}{%
{\protect \APACyear {2022}}%
}]{%
MPI_BMSB}
\APACinsertmetastar {%
MPI_BMSB}%
\begin{APACrefauthors}%
{Ministry for Primary Industries}.%
\end{APACrefauthors}%
\unskip\
\newblock
\APACrefYearMonthDay{2022}{{\APACmonth{05}}}{}.
\newblock
\APACrefbtitle {Winter campaign to raise awareness of brown marmorated stink bug.} {Winter campaign to raise awareness of brown marmorated stink bug.}
\newblock
\begin{APACrefURL} \url{https://www.mpi.govt.nz/news/media-releases/winter-campaign-to-raise-awareness-of-brown-marmorated-stink-bug/} \end{APACrefURL}
\PrintBackRefs{\CurrentBib}

\bibitem [\protect \citeauthoryear {%
Powell%
}{%
Powell%
}{%
{\protect \APACyear {2015}}%
}]{%
Powell2015}
\APACinsertmetastar {%
Powell2015}%
\begin{APACrefauthors}%
Powell, M\BPBI R.%
\end{APACrefauthors}%
\unskip\
\newblock
\APACrefYearMonthDay{2015}{}{}.
\newblock
{\BBOQ}\APACrefatitle {Risk-Based Sampling: I Don't Want to Weight in Vain} {Risk-based sampling: I don't want to weight in vain}.{\BBCQ}
\newblock
\APACjournalVolNumPages{Risk Analysis}{35}{12}{2172-2182}.
\newblock
\begin{APACrefDOI} 10.1111/risa.12415 \end{APACrefDOI}
\PrintBackRefs{\CurrentBib}

\bibitem [\protect \citeauthoryear {%
Robinson%
, Burgman%
\BCBL {}\ \BBA {} Cannon%
}{%
Robinson%
\ \protect \BOthers {.}}{%
{\protect \APACyear {2011}}%
}]{%
Robinson2011}
\APACinsertmetastar {%
Robinson2011}%
\begin{APACrefauthors}%
Robinson, A.%
, Burgman, M\BPBI A.%
\BCBL {}\ \BBA {} Cannon, R.%
\end{APACrefauthors}%
\unskip\
\newblock
\APACrefYearMonthDay{2011}{jun}{}.
\newblock
{\BBOQ}\APACrefatitle {Allocating surveillance resources to reduce ecological invasions: maximizing detections and information about the threat} {Allocating surveillance resources to reduce ecological invasions: maximizing detections and information about the threat}.{\BBCQ}
\newblock
\APACjournalVolNumPages{Ecological Applications}{21}{4}{1410--1417}.
\newblock
\begin{APACrefDOI} 10.1890/10-0195.1 \end{APACrefDOI}
\PrintBackRefs{\CurrentBib}

\bibitem [\protect \citeauthoryear {%
Saccaggi%
, Wilson%
, Robinson%
\BCBL {}\ \BBA {} Terblanche%
}{%
Saccaggi%
\ \protect \BOthers {.}}{%
{\protect \APACyear {2022}}%
}]{%
Saccaggi2022}
\APACinsertmetastar {%
Saccaggi2022}%
\begin{APACrefauthors}%
Saccaggi, D\BPBI L.%
, Wilson, J\BPBI R\BPBI U.%
, Robinson, A\BPBI P.%
\BCBL {}\ \BBA {} Terblanche, J\BPBI S.%
\end{APACrefauthors}%
\unskip\
\newblock
\APACrefYearMonthDay{2022}{}{}.
\newblock
{\BBOQ}\APACrefatitle {Arthropods on imported plant products: Volumes predict general trends while contextual details enhance predictive power} {Arthropods on imported plant products: Volumes predict general trends while contextual details enhance predictive power}.{\BBCQ}
\newblock
\APACjournalVolNumPages{Ecological Applications}{32}{3}{e2554}.
\newblock
\begin{APACrefDOI} 10.1002/eap.2554 \end{APACrefDOI}
\PrintBackRefs{\CurrentBib}

\bibitem [\protect \citeauthoryear {%
{Tony Cai}%
}{%
{Tony Cai}%
}{%
{\protect \APACyear {2005}}%
}]{%
cai2005}
\APACinsertmetastar {%
cai2005}%
\begin{APACrefauthors}%
{Tony Cai}, T.%
\end{APACrefauthors}%
\unskip\
\newblock
\APACrefYearMonthDay{2005}{}{}.
\newblock
{\BBOQ}\APACrefatitle {One-sided confidence intervals in discrete distributions} {One-sided confidence intervals in discrete distributions}.{\BBCQ}
\newblock
\APACjournalVolNumPages{Journal of Statistical Planning and Inference}{131}{1}{63-88}.
\PrintBackRefs{\CurrentBib}

\bibitem [\protect \citeauthoryear {%
Turner%
\ \protect \BOthers {.}}{%
Turner%
\ \protect \BOthers {.}}{%
{\protect \APACyear {2021}}%
}]{%
Turner2021}
\APACinsertmetastar {%
Turner2021}%
\begin{APACrefauthors}%
Turner, R\BPBI M.%
, Brockerhoff, E\BPBI G.%
, Bertelsmeier, C.%
, Blake, R\BPBI E.%
, Caton, B.%
, James, A.%
\BDBL {}Liebhold, A\BPBI M.%
\end{APACrefauthors}%
\unskip\
\newblock
\APACrefYearMonthDay{2021}{aug}{}.
\newblock
{\BBOQ}\APACrefatitle {Worldwide border interceptions provide a window into human-mediated global insect movement} {Worldwide border interceptions provide a window into human-mediated global insect movement}.{\BBCQ}
\newblock
\APACjournalVolNumPages{Ecological Applications}{31}{7}{}.
\newblock
\begin{APACrefDOI} 10.1002/eap.2412 \end{APACrefDOI}
\PrintBackRefs{\CurrentBib}

\bibitem [\protect \citeauthoryear {%
Turner%
\ \protect \BOthers {.}}{%
Turner%
\ \protect \BOthers {.}}{%
{\protect \APACyear {2020}}%
}]{%
Turner2020}
\APACinsertmetastar {%
Turner2020}%
\begin{APACrefauthors}%
Turner, R\BPBI M.%
, Plank, M\BPBI J.%
, Brockerhoff, E\BPBI G.%
, Pawson, S.%
, Liebhold, A.%
\BCBL {}\ \BBA {} James, A.%
\end{APACrefauthors}%
\unskip\
\newblock
\APACrefYearMonthDay{2020}{}{}.
\newblock
{\BBOQ}\APACrefatitle {Considering unseen arrivals in predictions of establishment risk based on border biosecurity interceptions} {Considering unseen arrivals in predictions of establishment risk based on border biosecurity interceptions}.{\BBCQ}
\newblock
\APACjournalVolNumPages{Ecological Applications}{30}{8}{e02194}.
\newblock
\begin{APACrefDOI} 10.1002/eap.2194 \end{APACrefDOI}
\PrintBackRefs{\CurrentBib}

\bibitem [\protect \citeauthoryear {%
{United Nations Conference on Trade and Developement}%
}{%
{United Nations Conference on Trade and Developement}%
}{%
{\protect \APACyear {2022}}%
}]{%
UNCTAD_stats_2022}
\APACinsertmetastar {%
UNCTAD_stats_2022}%
\begin{APACrefauthors}%
{United Nations Conference on Trade and Developement}.%
\end{APACrefauthors}%
\unskip\
\newblock
\APACrefYearMonthDay{2022}{}{}.
\newblock
\APACrefbtitle {{UNCTAD Handbook of Statistics 2022}.} {{UNCTAD Handbook of Statistics 2022}.}
\newblock
\begin{APACrefURL} \url{https://unctad.org/system/files/official-document/tdstat47_en.pdf} \end{APACrefURL}
\PrintBackRefs{\CurrentBib}

\bibitem [\protect \citeauthoryear {%
Whyte%
}{%
Whyte%
}{%
{\protect \APACyear {2009}}%
}]{%
Whyte2009_ISPM}
\APACinsertmetastar {%
Whyte2009_ISPM}%
\begin{APACrefauthors}%
Whyte, C\BPBI F.%
\end{APACrefauthors}%
\unskip\
\newblock
\APACrefYearMonthDay{2009}{}{}.
\newblock
\APACrefbtitle {{Explanatory Document on ISPM 31 Methodologies for sampling of consignments}} {{Explanatory Document on ISPM 31 Methodologies for sampling of consignments}}\ \APACbVolEdTR{}{}.
\newblock
\APACaddressInstitution{}{IPPC}.
\newblock
\begin{APACrefURL} \url{https://www.ippc.int/en/publications/43/} \end{APACrefURL}
\PrintBackRefs{\CurrentBib}

\bibitem [\protect \citeauthoryear {%
{World Trade Organization}%
}{%
{World Trade Organization}%
}{%
{\protect \APACyear {1994}}%
}]{%
WTOALOP}
\APACinsertmetastar {%
WTOALOP}%
\begin{APACrefauthors}%
{World Trade Organization}.%
\end{APACrefauthors}%
\unskip\
\newblock
\APACrefYearMonthDay{1994}{}{}.
\newblock
\APACrefbtitle {{Agreement on the Application of Sanitary and Phytosanitary Measures}.} {{Agreement on the Application of Sanitary and Phytosanitary Measures}.}
\newblock
\begin{APACrefURL} \url{https://www.wto.org/english/docs_e/legal_e/15sps_01_e.htm} \end{APACrefURL}
\PrintBackRefs{\CurrentBib}

\bibitem [\protect \citeauthoryear {%
{World Trade Organization}%
}{%
{World Trade Organization}%
}{%
{\protect \APACyear {2023}}%
}]{%
WTO_stats}
\APACinsertmetastar {%
WTO_stats}%
\begin{APACrefauthors}%
{World Trade Organization}.%
\end{APACrefauthors}%
\unskip\
\newblock
\APACrefYearMonthDay{2023}{}{}.
\newblock
\APACrefbtitle {Evolution of trade under the WTO: handy statistics.} {Evolution of trade under the wto: handy statistics.}
\newblock
\begin{APACrefURL} \url{https://www.wto.org/english/res_e/statis_e/trade_evolution_e/evolution_trade_wto_e.htm} \end{APACrefURL}
\newblock
\APACrefnote{Accessed 2023.}
\PrintBackRefs{\CurrentBib}

\end{thebibliography}

\appendix

\bmsection{Arrangement for numerical calculation}\label{app:technical}
\vspace*{12pt}

To calculate the integral from Equation~\eqref{eq:prob_correct_N1}, we need the conditional truncated Beta distribution $\mathbb{P}\left(r\left| r>T_{\text{risk}},N_0, y_0\right.\right)$. To be able to calculate this more easily numerically, we make some transformations. Note that if the prior rate is significantly smaller than $T_{\text{risk}}$, then $\mathbb{P}\left(r\left|N_0, y_0\right.\right)\approx0$ for $r>T_{\text{risk}}$. As such, it is numerically more precise to calculate: 
\begin{align}
& \log \mathbb{P}\left(r\left| r>T_{\text{risk}},N_0, y_0\right.\right)\nonumber\\
& =(\alpha_{0}-1)\log(r)+(\beta_{0}-1)\log(1-r)-\log(M),
\end{align}
where  $\alpha_{0}=y_{0}+0.5$, $\beta_{0}=N_{0}-y_{0}+0.5$, and 
where the normalising factor $M$ is: 
\begin{align}
M = \int_{0}^{-\log T_{\text{risk}}}\mathrm{d}t\,e^{-\alpha_{0}t}(1-e^{-t})^{\beta_{0}-1},
\end{align}
with the change of variables $r=e^{-t}$ in the integral to improve numerical calculation precision.

With this, we rewrite the probability $\mathbb{P}\left(r\left| r>T_{\text{risk}},N_0, y_0\right.\right)$ as:
\begin{align}
&\mathbb{P}\left(r\left| r>T_{\text{risk}},N_0, y_0\right.\right) \nonumber\\
& =\dfrac{\mathbb{P}\left(r\left| N_0, y_0\right.\right)}{\int_{T_{\text{risk}}}^{1}dr^\prime\mathbb{P}\left(r^\prime\left| N_0, y_0\right.\right)},\qquad \text{for }r>T_\text{risk} \\ 
 & =\exp\left[ \begin{array}{l}
\log \mathbb{P}\left(r\left| N_0, y_0\right.\right) \\
- \log\left(\int_{T_{\text{risk}}}^{1}dr^{\prime}\mathbb{P}\left(r^\prime\left| N_0, y_0\right.\right)\right)\\
\end{array} \right].
\end{align}
Then, using the definition of a Beta distribution $\mathbb{P}\left(r\left| N_0, y_0\right.\right)$, with beta function $\text{Beta}(\alpha_{0},\beta_{0})$, 
\begin{align}
\log & \mathbb{P}\left(r\left| N_0, y_0\right.\right) \nonumber \\
&=\log\left(r^{\alpha_{0}-1}(1-r)^{\beta_{0}-1}\right)-\log \text{Beta}(\alpha_{0},\beta_{0}),
\end{align}
and,
\begin{align}
&\log\left(\int_{T_{\text{risk}}}^{1}dr^{\prime}\mathbb{P}\left(r^{\prime}\left|N_0, y_0,\right.\right)\right)\nonumber\\ 
& =\log\left(\int_{T_{\text{risk}}}^{1}dr^{\prime}((r^{\prime})^{\alpha_{0}-1}(1-r^{\prime})^{\beta_{0}-1}\right)\\
&\qquad -\log \text{Beta}(\alpha_{0},\beta_{0}).
\end{align}
Additionally, we can make the change of variables  $r^\prime =e^{-t}$ in the integral. Hence, the conditional truncated Beta distribution can also be written as:
\begin{align}
& \mathbb{P}\left(r\mid r>T_{\text{risk}} \left| N_0, y_0 \right.\right)\nonumber\\
& =\exp\left[\begin{array}{r}
(\alpha_{0}-1)\log(r)+(\beta_{0}-1)\log(1-r)\\
-\int_{0}^{-\log T_{\text{risk}}}\mathrm{d}t\,e^{-\alpha_{0}t}(1-e^{-t})^{\beta_{0}-1}
\end{array}\right],
\end{align}
which is the form we use as part of the numerical sample size calculation.

\bmsection{Normal approximation to the binomial}\label{app:Normal-approximation}
\vspace*{12pt}

To speed up calculations, we introduced a normal approximation to the various binomials that appears in the numerical calculations. 

Consider the binomial random variable 
\begin{align}
y & \sim \text{Binomial}(N,r),
\end{align}
where $y$ is the number of leakages, $N$ is the sample size, and $r$ is the probability of a leakage. The binomial can be approximated
by the normal distribution with mean $Nr$ and standard deviation $\sqrt{N\cdot r\cdot(1-r)}$, i.e, 
\begin{align}
\mathcal{N}\left(N\cdot r,\sqrt{N\cdot r\cdot(1-r)}\right) & .
\end{align}

\end{document}